\documentclass[english,notitlepage,pra,twocolumn]{revtex4-1}
\usepackage[T1]{fontenc}
\usepackage[latin9]{inputenc}
\usepackage{amsmath}
\usepackage{amssymb}
\usepackage{graphicx}

\makeatletter
\usepackage[colorlinks,urlcolor=cyan,citecolor=blue,linkcolor=magenta]{hyperref}

\makeatother

\usepackage{babel}
\begin{document}
\title{Dynamic virial theorem at nonequilibrium and applications}
\author{Shi-Guo Peng}
\email{pengshiguo@wipm.ac.cn}

\affiliation{State Key Laboratory of Magnetic Resonance and Atomic and Molecular
Physics, Innovation Academy for Precision Measurement Science and
Technology, Chinese Academy of Sciences, Wuhan 430071, China}
\date{\today}
\begin{abstract}
We show that a variety of nonequilibrium dynamics of interacting many-body
systems are universally characterized by an elegant relation, which
we call the dynamic virial theorem. The out-of-equilibrium dynamics
of quantum correlations is entirely governed by Tan\textquoteright s
contact. It gives rise to a series of observable consequences and
is closely related to experiments with ultracold atoms. Especially,
we show that the dynamic virial theorem provides an experimentally
accessible verification of maximum energy growth theorem {[}Qi \emph{et
al.}, Phys. Rev. Lett. 126, 240401 (2021){]}, which is encoded in
the evolution of the atomic cloud size during expansion. In addition,
the dynamic virial theorem leads to a simple thermodynamic relation
of strongly interacting quantum gases in the framework of two-fluid
hydrodynamic theory, which holds in a wide range of temperature. This
thermodynamic relation is a kind of the out-of-equilibrium analog
of Tan's pressure relation at equilibrium. Our results provide fundamental
understanding of generic behaviors of interacting many-body systems
at nonequilibrium, and are readily examined in experiments with ultracold
atoms.
\end{abstract}
\maketitle

\section{Introduction}

Investigating nonequilibrium dynamics of strongly-interacting many-body
systems is of fundamental importance and remains an open challenge
in modern physics, while it is broadly relevant to many phenomena
in universe, ranging from the evolution of neutron stars, procedure
of chemical reactions to complex living systems in biology. Discovering
universal features in quantum systems at nonequilibrium, irrelevant
to the microscopic detail of objects studied, is a long-standing challenge
to date and attracts a great deal of attention in both theory and
experiment. Ultracold atoms, as a clean, controllable and versatile
quantum systems, provide an unprecedented platform for exploration
of a wide variety of phenomena in many-body physics, ranging from
thermodynamic equilibrium to out-of-equilibrium dynamics \citep{Bloch2008M,Giorgini2008T,Bloch2012Q,Polkovnikov2011C,Langen2015U,Eisert2015Q}.
Remarkably, ultracold atomic systems take unique advantages for studying
nonequilibrium dynamics within experimentally resolved intrinsic time
scales (typically milliseconds), holding promising opportunities to
test universality of many-body dynamics far from equilibrium \citep{Makotyn2014U,Eigen2017U,Eigen2018U,Prufer2018O,Erne2018U,Ko2019K,Dyke2021D}. 

A set of universal relations in cold atoms, connected by a simple
contact parameter, have been discovered \citep{Tan2008E,Tan2008L,Tan2008G}.
These relations simply follow from the short-range correlation of
two-body physics and provide remarkable understanding of profound
properties of interacting many-body systems. To date, an impressive
amount of experimental and theoretical efforts have been devoted to
confirm universal relations and explore their important consequences
\citep{Zwerger2011,Yu2015U,He2016C,Luciuk2016E,Peng2016L,Fletcher2017Q,Peng2018C,Zhang2020U}.
While most of them are focused on properties of many-body systems
at equilibrium, universal relations at nonequilibrium are rarely studied
and still remain elusive.

In this work, it is found that a variety of dynamic processes of many-body
systems driven away from equilibrium by either the time-dependent
external potential or time-dependent interactions are elegantly governed
by the dynamic virial theorem. These are favorite ways to study out-of-equilibrium
dynamics in experiments of cold atoms, such as the expansion dynamics
\citep{Elliott2014O,Deng2016O} and the interaction quench \citep{Makotyn2014U,Eigen2017U,Eigen2018U,Prufer2018O,Dyke2021D}.
The dynamic virial theorem reveals a deep insight into the precise
energy conversion relation, i.e.,
\begin{equation}
E\left(t\right)-2E_{ho}\left(t\right)=\frac{1}{4}\frac{d^{2}I\left(t\right)}{dt^{2}}-\frac{\hbar^{2}\mathcal{C}\left(t\right)}{8\pi ma\left(t\right)},\label{eq:DynamicalVT}
\end{equation}
that imposes an intrinsic constrain on energy dynamics. Here, $m$
is the atomic mass, $\hbar$ is Planck's constant, and $a\left(t\right)$
is the scattering length generally varying in time. $E_{ho}\left(t\right)$
is the energy corresponding to the external harmonic potential, $I\left(t\right)\equiv\left\langle mr^{2}\right\rangle $
is the moment of inertia of systems, and $\mathcal{C}\left(t\right)$
is Tan's contact \citep{Tan2008E,Tan2008L,Tan2008G}. In the classical
(or high-temperature) limit, the kinetic energy dominates the internal
energy. The dynamic virial theorem immediately implies an interesting
non-damping monopole oscillation with twice the trapping frequency
in a perfectly spherical trap, and systems never reach thermal equilibrium.
This long-predicted phenomenon by Boltzmann's equation has recently
been observed by JILA's group in a thermal Bose gas \citep{Lobser2015O}.
In the quantum (or low-temperature) limit, the interaction between
particles comes into play, and shifts the frequency of monopole oscillation
\citep{Stringari1996C,Bruun1999H,Baranov2000L,Guey-Odelin2002M,SupplementalMaterial}.
We find that the dynamics of quantum correlations is entirely governed
by Tan's contact. In the unitarity limit with divergent scattering
length, the systems display scale invariance. The dynamic virial theorem
requires a unitary gas to follow exactly the same dynamics of an ideal
gas \citep{Elliott2014O,Deng2016O,Werner2006U}. At equilibrium, the
dynamic virial theorem simply recovers that of \citep{Tan2008G} at
finite interaction strength, and further reduces to the well-known
equilibrium result of $E=2E_{ho}$ for unitary and ideal gases \citep{Thomas2005V}. 

The applications of the dynamic virial theorem are demonstrated in
several typical dynamic processes. The observable consequences in
experiments of ultracold atoms are discussed. There are two typical
ways to study the nonequilibirum dynamics in experiments with cold
atoms, i.e., in a time-dependent trap and with a time-dependent interaction.
For the former, we show that the release energy of cold atoms during
free expansion is directly given by the dynamic virial theorem. The
frequencies of monopole oscillations are simply derived according
to the dynamic virial theorem. Besides, the dynamic virial theorem
simply characterizes the intriguing Efimovian expansion of scale-invariant
quantum gases in a time-dependent harmonic trap \citep{Deng2016O}.
For the latter, a fundamental issue has theoretically been addressed
recently as how fast the energy could be pumped into a noninteracting
system by increasing interactions \citep{Qi2021M}. Counterintuitively,
it is not true that the faster the interaction increases, the larger
the rate of energy gain becomes. Instead, there exists an upper limit
of initial energy growth rate, which could be reached only when the
scattering length increases with time $t$ as $\sim\sqrt{t}$, known
as the maximum energy growth theorem. However, the direct measurement
of the energy dynamics in experiments is difficult. We show that the
dynamic virial theorem provides an experimentally accessible way to
verify the maximum energy growth theorem according to the expansion
behavior of the atomic cloud size. Remarkably, a simple dynamic thermodynamic
relation is derived by using the dynamic virial theorem in the framework
of the two-fluid hydrodynamic theory for interacting quantum gases.
This thermodynamic relation is a kind of the out-of-equilibrium analog
of Tan's pressure relation at equilibrium \citep{Tan2008G}, and holds
in a wide range of temperature.

The rest of the paper is arranged as follows. In the next section,
we present the brief proof of the dynamic virial theorem for interacting
many-body systems. Afterwards, the dynamic virial theorem is applied
to study the out-of-equilibrium dynamics for cold atoms in time-dependent
harmonic traps, i.e., free expansion (Sec. \ref{sec:Free-expansion-of}),
Efimovian expansion (Sec. \ref{sec:Efimovian-expansion}), and monopole
oscillation (Sec. \ref{sec:Monopole-oscillations}). We show, in Sec.
\ref{sec:Maximum-energy-growth}, that the dynamic virial theorem
provides an experimentally accessible verification of the maximum
energy growth theorem by increasing interatomic interactions, which
is encoded in the evolution of the atomic cloud size during expansion.
An out-of-equilibrium thermodynamic relations is obtained by using
the dynamic virial theorem in the framework of the two-fluid hydrodynamic
theory in Sec. \ref{sec:Thermodynamic-relation}. Finally, the remarks
and conclusions are summarized in Sec. \ref{sec:Conclusions}.

\section{dynamic virial theorem}

To prove the dynamic virial theorem (\ref{eq:DynamicalVT}), let us
consider a system consisting of $N$ atoms (either bosons or fermions)
in a harmonic trap. The Hamiltonian of the system takes the form of
\begin{equation}
\hat{H}=\hat{H}_{0}+\hat{V}_{int},\label{eq:2.1}
\end{equation}
in which 
\begin{equation}
\hat{H}_{0}=\sum_{j=1}^{N}\left[{\bf p}_{j}^{2}/2m+\hat{V}_{ho}\left({\bf r}_{j};t\right)\right]\label{eq:2.2}
\end{equation}
is the single-particle Hamiltonian, and 
\begin{equation}
\hat{V}_{int}=\sum_{ij}\hat{U}\left({\bf r}_{ij};a\left(t\right)\right)\label{eq:2.3}
\end{equation}
is the interatomic interactions with ${\bf r}_{ij}={\bf r}_{i}-{\bf r}_{j}$.
Here, we consider the most general case that the harmonic potential
$\hat{V}_{ho}$ and interatomic interactions both vary in time, i.e.,
with the time-dependent trapping frequency $\omega\left(t\right)$
and $s$-wave scattering length $a\left(t\right)$. The evolution
of moment of inertia $I\left(t\right)=\left\langle \sum_{j}mr_{j}^{2}\right\rangle \equiv\left\langle mr^{2}\right\rangle $
is governed by Heisenberg equation, i.e., 
\begin{equation}
\frac{dI}{dt}=\frac{1}{i\hbar}\left[I,\hat{H}\right],\label{eq:2.4}
\end{equation}
which simply yields $\dot{I}\equiv dI/dt=2\left\langle \hat{D}\right\rangle $,
and $\hat{D}=\sum_{j=1}^{N}\left({\bf r}_{j}\cdot{\bf p}_{j}+{\bf p}_{j}\cdot{\bf r}_{j}\right)/2$
is the generator of scale transformation named the dilatation operator
\citep{Pitaevskii1997B,Werner2006U,Maki2019Q}. By further taking
the second-order derivative with respect to time, i.e., $\ddot{I}=2\left\langle \left[\hat{D},\hat{H}\right]\right\rangle /i\hbar$,
we obtain
\begin{equation}
\frac{d^{2}I}{dt^{2}}=4\left\langle \hat{H}\right\rangle -8\left\langle \hat{V}_{ho}\right\rangle -4\left\langle \hat{V}_{int}\right\rangle +\frac{2}{i\hbar}\left\langle \left[\hat{D},\hat{V}_{int}\right]\right\rangle .\label{eq:2.5}
\end{equation}
To proceed, we notice that the two-body interaction $\hat{U}\left({\bf r}_{ij};a\right)$,
under the scale transformation, has the property of $e^{-i\epsilon\hat{D}/\hbar}\hat{U}\left({\bf r}_{ij};a\right)e^{i\epsilon\hat{D}/\hbar}=e^{2\epsilon}\hat{U}\left({\bf r}_{ij};e^{\epsilon}a\right)$
for infinitesimal $\epsilon$. Expanding both sides of this equation
up to the first-order term of $\epsilon$, we obtain the commutation
relation of $\left[\hat{D},\hat{U}\right]$, which in turn gives
\begin{equation}
\frac{1}{i\hbar}\left\langle \left[\hat{D},\hat{V}_{int}\right]\right\rangle =2\left\langle \hat{V}_{int}\right\rangle +a\left\langle \frac{\partial\hat{V}_{int}}{\partial a}\right\rangle .\label{eq:2.6}
\end{equation}
Inserting Eq.(\ref{eq:2.6}) into Eq.(\ref{eq:2.5}) and combining
with Hellmann-Feynman theorem and Tan's relation \citep{Tan2008L},
i.e.,
\begin{equation}
\left\langle \frac{\partial\hat{V}_{int}}{\partial a}\right\rangle =\frac{\partial E}{\partial a}=\frac{\hbar^{2}\mathcal{C}}{4\pi ma^{2}},\label{eq:2.7}
\end{equation}
we finally arrive at Eq.(\ref{eq:DynamicalVT}) with $E\left(t\right)=\left\langle \hat{H}\right\rangle $
and $E_{ho}\left(t\right)=\left\langle \hat{V}_{ho}\right\rangle $.
The similar formulism has been obtained in discussing the breathing
mode of two-dimensional Fermi gases and corresponding scale-invariant
dynamics, such as the quantum anomaly \citep{Hofmann2012Q,Gao2012B}.
Here, we emphasize that Eq.(\ref{eq:DynamicalVT}), as a fundamental
analogous relation of virial theorem at equilibrium \citep{Thomas2005V},
characterizes a broad variety of nonequilibrium dynamics of interacting
quantum gases. It provides a deep insight into generic out-of-equilibrium
characters, which we are going to demonstrate as in the follows.

\section{Free expansion of quantum gases\label{sec:Free-expansion-of}}

Free expansion, simply releasing atoms from the trap, provides crucial
information on both equilibrium and dynamic properties of ultracold
atomic gases. It is widely used in experiments of cold atoms, such
as the time of flight. One of direct applications of dynamic virial
theorem is to calculate the release energy, i.e., $E_{rel}\equiv E\left(t\right)$.
Since the release energy is well understood in weakly-interacting
and unitary gases, it is inversely convenient to verify the validity
of dynamic virial theorem. To this end, the knowledge of evolution
of Tan's contact as well as that of moment of inertia is required.
As an example, let us consider free expansion of a two-component Fermi
gas, initially prepared at equilibrium in the trap. We evaluate the
evolution of Tan's contact and moment of inertia of the system in
the framework of hydrodynamic theory \citep{Menotti2002E,Hu2004C}.
To solve the hydrodynamic equations, we adopt the scaling form of
time-dependent density profile, i.e., $n\left({\bf r},t\right)=n_{0}\left({\bf r}/b\right)/b^{3}\left(t\right)$,
where $n_{0}\left({\bf r}\right)$ is the equilibrium density profile
initially in the trap. The time dependence of $n\left({\bf r},t\right)$
is entirely determined by the scaling factor $b\left(t\right)$. The
atomic cloud size $\left\langle r^{2}\right\rangle \left(t\right)$
during expansion is related to the initial size $\left\langle r^{2}\right\rangle _{0}$
by the scaling factor $b\left(t\right)$ as $\left\langle r^{2}\right\rangle \left(t\right)=b^{2}\left(t\right)\left\langle r^{2}\right\rangle _{0}$,
which in turn gives 
\begin{equation}
\ddot{I}\left(t\right)=\frac{4}{\omega_{0}}\left[\dot{b}^{2}\left(t\right)+b\left(t\right)\ddot{b}\left(t\right)\right]E_{ho}^{(0)},\label{eq:3.1}
\end{equation}
$E_{ho}^{(0)}$ is the initial potential energy and $\omega_{0}$
is the initial trapping frequency. 

In the weakly-interacting Bardeen-Cooper-Schrieffer (BCS) limit with
small negative scattering length, the scaling factor $b\left(t\right)$
satisfies \citep{Menotti2002E,Hu2004C}
\begin{equation}
\ddot{b}\left(t\right)-\frac{\omega_{0}^{2}}{b^{3}\left(t\right)}+\frac{3}{2}\chi\left(g\right)\omega_{0}^{2}\left[\frac{1}{b^{3}\left(t\right)}-\frac{1}{b^{4}\left(t\right)}\right]=0,\label{eq:3.2}
\end{equation}
 where $\chi\left(g\right)=E_{int}^{(0)}/E_{ho}^{(0)}$ is the ratio
of interaction energy $E_{int}^{(0)}$ to potential energy $E_{ho}^{(0)}$
initially in the trap, and $g=4\pi\hbar^{2}a/m$ is the interaction
strength between atoms. The contact $\mathcal{C}\left(t\right)$ during
expansion may be evaluated according to the local density approximation
(LDA) by making the adiabatic ansatz \citep{Qu2016E}, i.e., $\mathcal{C}\left(t\right)=\int d{\bf r}\mathcal{I}\left({\bf r},t\right)$
with local contact density $\mathcal{I}\left({\bf r},t\right)\approx4\pi^{2}n^{2}\left({\bf r},t\right)a^{2}$
\citep{Tan2008L}, which in turn gives $\mathcal{C}\left(t\right)=4\pi maE_{int}^{(0)}/\hbar^{2}b^{3}$.
With all these results in hands, we easily obtain the release energy
according to the dynamic virial theorem, i.e., 
\begin{equation}
E_{rel}=E_{ho}^{(0)}\left\{ \frac{1}{\omega_{0}^{2}}\left[\dot{b}^{2}\left(t\right)+b\left(t\right)\ddot{b}\left(t\right)\right]-\frac{\chi\left(g\right)}{2b^{3}\left(t\right)}\right\} \label{eq:3.3}
\end{equation}
From Eq.(\ref{eq:3.2}), we find \citep{Derivation1}
\begin{equation}
\frac{1}{\omega_{0}^{2}}\left[\dot{b}^{2}\left(t\right)+b\left(t\right)\ddot{b}\left(t\right)\right]=1-\frac{1}{2}\chi\left(g\right)\left[1-\frac{1}{b^{3}\left(t\right)}\right].\label{eq:3.4}
\end{equation}
Then the release energy reads 
\begin{equation}
E_{rel}=\left[1-\frac{1}{2}\chi\left(g\right)\right]E_{ho}^{(0)}\label{eq:3.5}
\end{equation}
 in the weakly-interacting limit, and further reduces to the well-known
result $E_{rel}=E_{ho}^{(0)}$ in the BCS limit \citep{Giorgini2008T}.

In the Bose-Einstein-Condensation (BEC) limit with small positive
scattering length, the scaling factor satisfies \citep{Hu2004C}
\begin{equation}
\ddot{b}\left(t\right)-\frac{\omega_{0}^{2}}{b^{4}\left(t\right)}=0.\label{eq:3.6}
\end{equation}
It is nothing than that of weakly-interacting bosons \citep{Castin1996B,Kagan1996E}.
The evolution of contact is again determined by $\mathcal{C}\left(t\right)=\int d{\bf r}\mathcal{I}\left({\bf r},t\right)$
with local contact density \citep{Tan2008L}
\begin{equation}
\mathcal{I}\left({\bf r},t\right)\approx\frac{4\pi n\left({\bf r},t\right)}{a}+\pi^{2}n^{2}\left({\bf r},t\right)a_{m}a,\label{eq:3.7}
\end{equation}
where $a_{m}\approx0.6a$ is the scattering length between molecules
\citep{Petrov2004}. Then we have $\mathcal{C}\left(t\right)=\mathcal{C}_{b}+\mathcal{C}_{m}b^{-3}$,
in which $\mathcal{C}_{b}=4\pi N/a$ is the contribution from the
binding energy of tightly bound molecules that is not released during
expansion. $\mathcal{C}_{m}=\int d{\bf r}\pi^{2}n_{0}^{2}\left({\bf r}\right)a_{m}a$
is the initial contact corresponding to the interaction energy between
molecules, i.e., $E_{m,int}^{(0)}=\hbar^{2}\mathcal{C}_{m}/4\pi ma$.
From the dynamic virial theorem, the release energy is easily calculated,
i.e., 
\begin{multline}
E_{rel}\equiv E+\frac{N\epsilon_{b}}{2}\\
=\left[\frac{1}{3b^{3}\left(t\right)}+\frac{2}{3}\right]E_{ho}^{(0)}-\frac{1}{2b^{3}\left(t\right)}E_{m,int}^{(0)},\label{eq:3.8}
\end{multline}
where $\epsilon_{b}=\hbar^{2}/ma^{2}$ is the binding energy of molecules.
By further using the equilibrium virial theorem for weakly-interacting
bosons initially in the trap $E_{m,int}^{(0)}=2E_{ho}^{(0)}/3$ \citep{Pitaevskii2016B},
the release energy simply becomes $E_{rel}=2E_{ho}^{(0)}/3$, the
well-known result in the BEC limit \citep{Giorgini2008T}. 

In the unitarity limit with divergent scattering length, the release
energy is rather easy to calculate, since the system obeys an exact
scale-invariant evolution, which yields $b\left(t\right)=\sqrt{1+\omega_{0}^{2}t^{2}}$
\citep{Werner2006U,Maki2019Q}. Besides, the second term on the right-hand
side of Eq.(\ref{eq:DynamicalVT}) vanishes in the unitarity limit.
Consequently, the dynamic virial theorem simply gives the well-known
result $E_{rel}=E_{ho}^{(0)}$ \citep{Giorgini2008T}.

One another application of free expansion is to test the scale invariance
of strongly-interacting quantum gases as well as important consequences
resulted from the scale-symmetry breaking \citep{Elliott2014O,Saint-Jalm2019D,Maki2022D}.
Such scale-invariant dynamics could be identified according to the
evolution of cloud size, which is governed by the dynamic virial theorem.
If defining 
\begin{equation}
\tau^{2}\left(t\right)\equiv\frac{m}{2E_{ho}^{(0)}}\left(\left\langle r^{2}\right\rangle -\left\langle r^{2}\right\rangle _{0}\right),\label{eq:3.9}
\end{equation}
we easily obtain the equation satisfied by $\tau^{2}\left(t\right)$
\begin{equation}
\frac{d^{2}}{dt^{2}}\tau^{2}\left(t\right)=2+\frac{\hbar^{2}\left[\mathcal{C}\left(t\right)-\mathcal{C}_{0}\right]}{4\pi maE_{ho}^{(0)}},\label{eq:3.10}
\end{equation}
where $\mathcal{C}_{0}$ is the initial contact before release. In
\citep{Elliott2014O}, $\tau^{2}\left(t\right)$, as an identification
of the scale invariance, is measured in the free expansion and obeys
$\tau^{2}\left(t\right)=t^{2}$ for a scale-invariant Fermi gas. In
the weakly-interacting limit, we obtain 
\begin{equation}
\frac{d^{2}}{dt^{2}}\tau^{2}\left(t\right)\approx2+0.29278k_{FI}a\left[\left(1+\omega_{0}^{2}t^{2}\right)^{-3/2}-1\right],\label{eq:3.11}
\end{equation}
and 
\begin{equation}
\frac{d^{2}}{dt^{2}}\tau^{2}\left(t\right)\approx2+2.4668\left(k_{FI}a\right)^{-1}\left[\left(1+\omega_{0}^{2}t^{2}\right)^{-1/2}-1\right]\label{eq:3.12}
\end{equation}
 in the strongly-interacting limit, where $k_{FI}$ is the Fermi wave
number at trap center for an ideal gas. Generally, an accurate estimation
of contact is needed to depict the evolution of $\tau^{2}\left(t\right)$
in the free expansion, for example, by using high-temperature virial
expansion \citep{Elliott2014O,Liu2009V,Liu2013V}.

\section{Efimovian expansion\label{sec:Efimovian-expansion}}

\begin{figure}
\includegraphics[width=0.8\columnwidth]{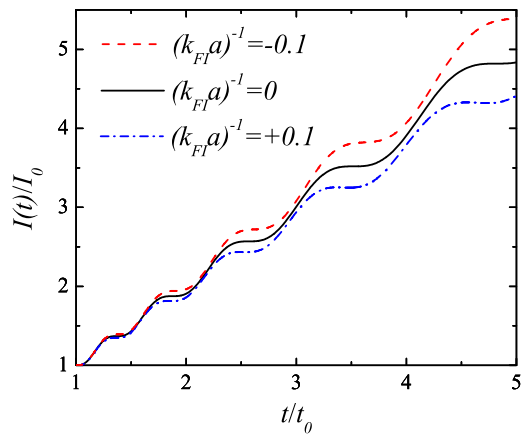}

\caption{(Color online) Efimovian expansion of a two-component Fermi gas near
resonance. Here, the variation rate of trap is set to be $\lambda=0.01$,
$t_{0}$ is the initial time, $I_{0}$ is the initial moment of inertia
(or cloud size), and $k_{FI}$ is the initial Fermi wave number at
trap center for an ideal gas.}

\label{fig:EfimovianExpansion}
\end{figure}

Scale-invariant quantum gases display a fancy scaling expansion dynamics
in time-dependent harmonic traps, i.e., $\omega\left(t\right)\sim1/\sqrt{\lambda}t$
with variation rate $\lambda$. This phenomenon is termed as ``Efimovian
expansion'' \citep{Deng2016O}. Here, we show that such profound
expansion dynamics is inherently governed by the dynamic virial theorem.
By further taking the derivative of dynamic virial theorem with respect
to $t$, we obtain
\begin{equation}
\frac{1}{4}\frac{d^{3}I}{dt^{3}}=\frac{dE}{dt}-2\frac{dE_{ho}}{dt}+\frac{\hbar^{2}}{8\pi ma}\frac{d\mathcal{C}}{dt}.\label{eq:4.1}
\end{equation}
According to Hellmann-Feynman theorem, we have
\begin{equation}
\frac{dE}{dt}=\frac{d}{dt}\left\langle \hat{H}\left(t\right)\right\rangle =\left\langle \frac{d}{dt}\hat{H}\left(t\right)\right\rangle =-\frac{I\left(t\right)}{\lambda t^{3}}.\label{eq:4.2}
\end{equation}
Combining with $E_{ho}=I/2\lambda t^{2}$, we arrive at
\begin{equation}
\frac{d^{3}I}{dt^{3}}+\frac{4}{\lambda t^{2}}\frac{dI}{dt}-\frac{4I}{\lambda t^{3}}=\frac{\hbar^{2}}{2\pi ma}\frac{d\mathcal{C}}{dt}.\label{eq:4.3}
\end{equation}
For noninteracting and resonant-interacting (scale-invariant) Fermi
gases, the right-hand side of Eq.(\ref{eq:4.3}) vanishes, which recovers
that of \citep{Deng2016O,MoI}. The evolution of cloud size demonstrates
a temporal scaling expansion behavior. Away from the scale-invariant
point at finite interaction strength, the expansion dynamics resulting
from the scale-symmetry breaking is simply characterized by the evolution
of contact \citep{Gharashi2016B,Qi2016T,Shi2016E}. In the strongly-interacting
limit, the contact can approximately be estimated as before, i.e.,
$\mathcal{C}\left(t\right)\approx\mathcal{C}_{0}/b\left(t\right)$
with $b^{2}\left(t\right)=I\left(t\right)/I_{0}$. Here, $I_{0}$
is the initial moment of inertia and $\mathcal{C}_{0}=256\pi\alpha Nk_{FI}/35\xi_{B}^{1/4}$
is the initial contact \citep{Qu2016E,Pitaevskii2016B}, where $N$
is the total atom number, $\alpha\approx0.12$ and $\xi_{B}\approx0.37$
are universal parameters. The evolution of cloud size (or moment of
inertia) during Efimovian expansion near resonant interaction is shown
in Fig.\ref{fig:EfimovianExpansion}.

\section{Monopole oscillations\label{sec:Monopole-oscillations}}

The study of low-energy elementary excitations is a subject of primary
importance in many-body physics. It is achieved, for example, by abruptly
disturbing the external trap and exerting the oscillation of system
around its equilibrium. In the follows, let us discuss the monopole
oscillation of cold atoms in a spherical trap, which we find is governed
by the dynamic virial theorem. 

Let us firstly consider a cluster of classical particles in a harmonic
trap interacting according to short-range potentials. The moment of
inertia of a single particle is defined as $I\equiv mr^{2}$. Then
we have $\dot{I}=2{\bf p}\cdot{\bf r}$ with the momentum ${\bf p}$.
By further taking the second-order derivative with respect to $t$,
we obtain $\ddot{I}=2\left(\dot{{\bf p}}\cdot{\bf r}+{\bf p}\cdot\dot{{\bf r}}\right)$.
As we have $\dot{{\bf p}}={\bf F}$, the force acting on the particle,
and the kinetic energy $E_{kin}={\bf p}\cdot\dot{{\bf r}}/2$, we
obtain $\ddot{I}=2\left({\bf F}\cdot{\bf r}+2E_{kin}\right)$. In
the harmonic trap, the force acting on the particle is the negative
gradient of the potential energy, i.e., ${\bf F}=-\nabla E_{ho}$,
which immediately gives ${\bf F}\cdot{\bf r}=-2E_{ho}$. Finally,
we obtain $E\left(t\right)-2E_{ho}\left(t\right)=\ddot{I}\left(t\right)/4$.
Since the interactions between atoms are local, the positions of atoms
as well as their moment of inertia are not changed from the instant
before to the instant after collision. Therefore, the momentum and
energy conversation implies that \citep{Lobser2015O}
\begin{equation}
E\left(t\right)-2E_{ho}\left(t\right)=\frac{1}{4}\frac{d^{2}I}{dt^{2}}\label{eq:5.1}
\end{equation}
holds for a cluster of classical particles. Eq.(\ref{eq:5.1}) is
the classical version of the dynamic virial theorem. The classical
dynamic virial theorem simply leads to a non-damping monopole oscillation
with the frequency $\omega_{M}=2\omega_{0}$, which has recently been
observed in a Bose gas \citep{Lobser2015O}.

In the low-temperature limit, quantum correlations resulting from
interactions come into play. Let us consider the small monopole oscillation
of a two-component Fermi gas initially prepared at equilibrium in
the trap. Then the total energy of the system is given by the virial
theorem, and we have $E=\left[2-\chi\left(g\right)/2\right]E_{ho}^{(0)}$
in the BCS limit, $E=2E_{ho}^{(0)}$ in the unitarity limit, and $E=5E_{ho}^{(0)}/3-N\epsilon_{b}/2$
in the BEC limit. Here, $\chi\left(g\right)=E_{int}^{(0)}/E_{ho}^{(0)}$
is the ratio of the interaction energy $E_{int}^{(0)}$ to the potential
energy $E_{ho}^{(0)}$, $\epsilon_{b}$ is the binding energy of molecules,
and $N$ is the atom number. At time $t=0$, the monopole oscillation
is exerted due to a slight disturbing on the trap. The evolution of
contact is easily evaluated by using the adiabatic ansatz as before.
Then the evolution of the moment of inertia or the cloud size is given
by the dynamic virial theorem. 

In the BCS limit, we obtain
\begin{equation}
\frac{d^{2}y}{dt^{2}}+4\omega_{0}^{2}y-\omega_{0}^{2}\chi\left(g\right)y^{-3/2}-2\omega_{0}^{2}\left[2-\frac{\chi\left(g\right)}{2}\right]=0,\label{eq:5.2}
\end{equation}
where we have defined $y\equiv\left\langle r^{2}\right\rangle \left(t\right)/\left\langle r^{2}\right\rangle \left(0\right)$.
By linearizing Eq.(\ref{eq:5.2}) around the equilibrium, i.e., $y\approx1+\delta y$,
we obtain
\begin{equation}
\frac{d^{2}\delta y}{dt^{2}}+\left[4+\frac{3}{2}\chi\left(g\right)\right]\omega_{0}^{2}\delta y=0.\label{eq:5.3}
\end{equation}
We find that the frequency of the monopole oscillation is shifted
to $\omega_{M}=\omega_{0}\sqrt{4+3\chi\left(g\right)/2}$ by interatomic
interactions. It further reduces to $\omega_{M}=2\omega_{0}$ in the
noninteracting limit. In the BEC limit, we similarly have
\begin{equation}
\frac{d^{2}y}{dt^{2}}+4\omega_{0}^{2}y-\frac{2}{3}\omega_{0}^{2}y^{-3/2}-\frac{10}{3}\omega_{0}^{2}=0,\label{eq:5.4}
\end{equation}
which is linearized to 
\begin{equation}
\frac{d^{2}\delta y}{dt^{2}}+5\omega_{0}^{2}\delta y=0.\label{eq:5.5}
\end{equation}
The frequency of the monopole oscillation is then shifted to $\omega_{M}=\sqrt{5}\omega_{0}$
in the BEC limit, consistent with the well-known result of Bose gases
in the presence of mean-field interactions. In the unitarity limit,
the dynamic virial theorem implies that a strongly interacting Fermi
gas follows exactly the same monopole oscillation as that of an ideal
gas with the frequency $\omega_{M}=2\omega_{0}$.

\section{Maximum energy growth under increasing interactions\label{sec:Maximum-energy-growth}}

In this section, let us discuss the dynamic behavior of interacting
many-body systems with time-dependent interatomic interactions. Explicitly,
we consider how fast the energy could be pumped into a system by increasing
interactions, starting from the non-interacting limit. This fundamental
issue has recently been addressed \citep{Qi2021M}. Counterintuitively,
it is not true that the faster the interaction increases, the larger
the rate of energy gain becomes. The maximum energy growth theorem
states that there exists an upper limit of initial energy growth rate,
which could be reached only when the scattering length increases with
time $t$ as $\sim\sqrt{t}$. However, the direct measurement of the
energy dynamics in experiments is difficult. Here, we are going to
show that the dynamic virial theorem provides an experimentally accessible
way to verify the maximum energy growth theorem according to the evolution
of the cloud size during expansion. 

Let us consider a two-component Fermi gas initially prepared at equilibrium
in the trap. It is supposed that the initial interatomic interaction
is so weak that we may treat it as an ideal gas. Then the gas is released
from the trap for $t\ge0$, accompanied by the increasing of interactions.
To estimate the \emph{initial} energy growth rate, it is sufficient
to consider the dynamics of the system in short time. Here, the short
time is defined as the timescale much smaller than the many-body timescale
$t_{n}=\hbar/E_{F}$ related to the Fermi energy $E_{F}$, and also
larger than the timescale $t_{r}=mr_{0}^{2}/\hbar$ imposed by the
range $r_{0}$ of the two-body potential. In this timescale, the scattering
length may be assumed to take a power-law form of 
\begin{equation}
a\left(t\right)=a_{ho}\beta\left(\omega_{0}t\right)^{\alpha},\label{eq:6.1}
\end{equation}
and the two-body physics dominates the expansion dynamics. Here, $a_{ho}=\sqrt{\hbar/m\omega_{0}}$
is the harmonic length with respect to the initial trap with frequency
$\omega_{0}$, and $\alpha,\beta>0$ are parameters characterize respectively
the power and quench rate. Then the energy can be obtained by simply
integrating Tan's sweep theorem over time $t$, i.e.,
\begin{equation}
E\left(t\right)=E\left(0\right)+\int_{0}^{t}dt^{\prime}\frac{\hbar^{2}\mathcal{C}\left(t^{\prime}\right)}{4\pi ma^{2}\left(t^{\prime}\right)}\frac{da}{dt^{\prime}},\label{eq:6.2}
\end{equation}
where $E\left(0\right)$ is the initial energy that is simply the
potential energy $E_{ho}\left(0\right)$ given by the virial theorem
at equilibrium. The evolution of the cloud size or the moment of inertia
is related to the energy characterized by the dynamic virial theorem
in expansion,
\begin{equation}
\frac{d^{2}I}{dt^{2}}=4E\left(t\right)+\frac{\hbar^{2}\mathcal{C}\left(t\right)}{2\pi ma\left(t\right)}.\label{eq:6.3}
\end{equation}
Inserting Eq.(\ref{eq:6.2}) into (\ref{eq:6.3}) , we obtain
\begin{equation}
\frac{d^{2}I}{dt^{2}}=2\omega_{0}^{2}I_{0}+\frac{\hbar^{2}\mathcal{C}\left(t\right)}{2\pi ma\left(t\right)}+\frac{\hbar^{2}}{\pi m}\int_{0}^{t}dt^{\prime}\frac{\mathcal{C}\left(t^{\prime}\right)}{a^{2}\left(t^{\prime}\right)}\frac{da}{dt^{\prime}},\label{eq:6.4}
\end{equation}
and $I_{0}=2E_{ho}\left(0\right)/\omega_{0}^{2}$ is the initial moment
of inertia. We find that the dynamics of Tan's contact is crucial
for evaluating the energy as well as the cloud size, which is simply
governed by the two-body physics in short time as pointed out in \citep{Qi2021M}.
There are totally three typical behaviors of the evolution of contact
in short time, depending on the power parameter $\alpha$ \citep{Qi2021M},
i.e., 
\begin{equation}
\mathcal{C}\left(t\right)=\begin{cases}
16\pi^{2}g_{2}\left(0\right)a^{2}\left(t\right), & \alpha>1/2,\\
\left|A\left(\beta\right)\right|^{2}g_{2}\left(0\right)\hbar t/m, & \alpha=1/2,\\
\left|A\left(\infty\right)\right|^{2}g_{2}\left(0\right)\hbar t/m, & 0<t<1/2,
\end{cases}\label{eq:6.5}
\end{equation}
where $g_{2}\left({\bf r}\right)$ is defined as 
\begin{equation}
g_{2}\left({\bf r}\right)\equiv\int d{\bf R}\rho_{2}\left({\bf R}+{\bf r}/2,{\bf R}-{\bf r}/2\right),
\end{equation}
and 
\begin{equation}
\rho_{2}\left({\bf r}_{1},{\bf r}_{2}\right)=\left\langle \hat{\psi}_{\uparrow}^{\dagger}\left({\bf r}_{1}\right)\hat{\psi}_{\downarrow}^{\dagger}\left({\bf r}_{2}\right)\hat{\psi}_{\downarrow}\left({\bf r}_{2}\right)\hat{\psi}_{\uparrow}\left({\bf r}_{1}\right)\right\rangle _{0}
\end{equation}
 is the two-body density matrix for the initial state. Here, it is
defined $A\left(\beta\right)=\left[B\left(1/2\right)+1/4\pi\beta\right]^{-1}$
and $B\left(\alpha\right)=i^{3/2}\Gamma\left(\alpha+1\right)/4\pi\Gamma\left(\alpha+1/2\right)$.
Subsequently, the energy as well as the evolution of the moment of
inertia can conveniently be evaluated as follows.

For $\alpha>1/2$, the contact takes the form of 
\begin{equation}
\mathcal{C}\left(t\right)=16\pi^{2}g_{2}\left(0\right)a^{2}\left(t\right)=16\pi^{2}g_{2}\left(0\right)a_{ho}^{2}\beta^{2}\left(\omega_{0}t\right)^{2\alpha}.\label{eq:6.6}
\end{equation}
The energy growth $\delta E\left(t\right)\equiv E\left(t\right)-E\left(0\right)$
in short time is obtained according to Eq.(\ref{eq:6.2}), 
\begin{equation}
\frac{\delta E\left(t\right)}{g_{2}\left(0\right)a_{ho}^{3}\hbar\omega_{0}}=4\pi\beta\left(\omega_{0}t\right)^{\alpha}.\label{eq:6.7}
\end{equation}
By inserting Eqs.(\ref{eq:6.1}) and (\ref{eq:6.6}) into Eq.(\ref{eq:6.4}),
the evolution of moment of inertia in short time satisfies
\begin{equation}
\frac{d^{2}I}{dt^{2}}=2\omega_{0}^{2}I_{0}+\frac{24\pi\hbar^{2}\beta}{m}g_{2}\left(0\right)a_{ho}\left(\omega_{0}t\right)^{\alpha}.\label{eq:6.8}
\end{equation}
Combining with the initial conditions $I\left(0\right)/I_{0}=1$,
and $\dot{I}\left(0\right)/I_{0}=0$, we obtain
\begin{equation}
f_{0}^{-1}\left[\frac{I}{I_{0}}-\left(1+\left(\omega_{0}t\right)^{2}\right)\right]=\frac{12\pi\beta}{\left(1+\alpha\right)\left(2+\alpha\right)}\left(\omega_{0}t\right)^{2+\alpha}\label{eq:6.9}
\end{equation}
with $f_{0}=g_{2}\left(0\right)a_{ho}^{3}\hbar\omega_{0}/E_{ho}\left(0\right)$.
If defining the reduced moment of inertia and energy growth as
\begin{eqnarray}
\tilde{I}\left(t\right) & \equiv & f_{0}^{-1}\left[\frac{I}{I_{0}}-\left(1+\left(\omega_{0}t\right)^{2}\right)\right],\label{eq:6.10}\\
\delta\tilde{E}\left(t\right) & \equiv & \frac{\delta E\left(t\right)}{\hbar\omega_{0}g_{2}\left(0\right)a_{ho}^{3}},\label{eq:6.11}
\end{eqnarray}
we easily find a one-on-one correspondence between the energy growth
and the moment of inertia
\begin{equation}
\delta\tilde{E}\left(t\right)=4\pi\beta\left[\frac{\left(1+\alpha\right)\left(2+\alpha\right)}{12\pi\beta}\tilde{I}\left(t\right)\right]^{\alpha/\left(2+\alpha\right)}.\label{eq:6.12}
\end{equation}

At the critical point of $\alpha=1/2$, we have
\begin{equation}
a\left(t\right)=\beta a_{ho}\left(\omega_{0}t\right)^{1/2}=\beta\left(\frac{\hbar}{m}\right)^{1/2}t^{1/2},\label{eq:6.13}
\end{equation}
and the contact is linearly dependent on time, i.e., $\mathcal{C}\left(t\right)=\left|A\left(\beta\right)\right|^{2}g_{2}\left(0\right)\hbar t/m$.
Then the reduced initial energy growth takes the form of
\begin{equation}
\delta\tilde{E}\left(t\right)=\frac{\left|A\left(\beta\right)\right|^{2}\left(\omega_{0}t\right)^{1/2}}{4\pi\beta}.\label{eq:6.14}
\end{equation}
Similarly, according to the dynamic virial theorem, the evolution
of the moment of inertia satisfies
\begin{equation}
\frac{d^{2}I}{dt^{2}}=2\omega_{0}^{2}I_{0}+\frac{3\hbar^{2}}{2\pi m\beta}\left|A\left(\beta\right)\right|^{2}g_{2}\left(0\right)a_{ho}\left(\omega_{0}t\right)^{1/2},\label{eq:6.15}
\end{equation}
which yields
\begin{equation}
\tilde{I}\left(t\right)=\frac{\left|A\left(\beta\right)\right|^{2}\left(\omega_{0}t\right)^{5/2}}{5\pi\beta}.\label{eq:6.16}
\end{equation}
Then we arrive at the one-on-one correspondence between the reduced
initial energy growth and the moment of inertia
\begin{equation}
\delta\tilde{E}\left(t\right)=\frac{\left|A\left(\beta\right)\right|^{8/5}}{4\pi\beta}\left[5\pi\beta\tilde{I}\left(t\right)\right]^{1/5}.\label{eq:6.17}
\end{equation}

For $0<\alpha<1/2$, the evolution of the contact linearly increases
with time as well, i.e., $\mathcal{C}\left(t\right)=\left|A\left(\infty\right)\right|^{2}g_{2}\left(0\right)\hbar t/m$,
and is independent of the parameters $\alpha$ and $\beta$. This
means that the contact is independent of how fast the scattering length
varies in time, even if the scattering length initially grows infinitely
fast as in the quench process. In this case, the corresponding reduced
energy growth becomes
\begin{equation}
\delta\tilde{E}\left(t\right)=\frac{16\alpha}{\beta\left(1-\alpha\right)}\left(\omega_{0}t\right)^{1-\alpha}.\label{eq:6.18}
\end{equation}
Again the evolution of the moment of inertia is given by the dynamic
virial theorem
\begin{equation}
\frac{d^{2}I}{dt^{2}}=2\omega_{0}^{2}I_{0}+\frac{32\hbar^{2}a_{ho}}{m\beta}\frac{1+\alpha}{1-\alpha}g_{2}\left(0\right)\left(\omega_{0}t\right)^{1-\alpha},\label{eq:6.19}
\end{equation}
which yields
\begin{equation}
\tilde{I}\left(t\right)=\frac{16\left(1+\alpha\right)}{\beta\left(1-\alpha\right)\left(2-\alpha\right)\left(3-\alpha\right)}\left(\omega_{0}t\right)^{3-\alpha}.\label{eq:6.20}
\end{equation}
Then we arrive at the relation
\begin{multline}
\delta\tilde{E}\left(t\right)=\frac{16\alpha}{\beta\left(1-\alpha\right)}\\
\times\left[\frac{\beta\left(1-\alpha\right)\left(2-\alpha\right)\left(3-\alpha\right)}{16\left(1+\alpha\right)}\tilde{I}\left(t\right)\right]^{\left(1-\alpha\right)/\left(3-\alpha\right)}.\label{eq:6.21}
\end{multline}

\begin{figure}
\includegraphics[width=1\columnwidth]{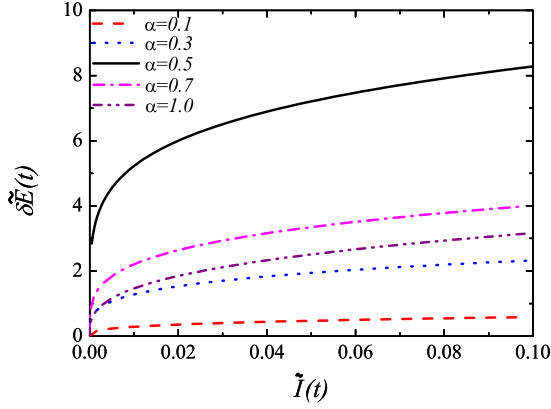}

\caption{(Color online) The initial energy growth as a function of the atomic
cloud size during the expansion accompanied by an increasing of the
scattering length $a\left(t\right)/a_{ho}=\beta\left(\omega_{0}t\right){}^{\alpha}$
at different power parameters. Here, we set $\beta=1$.}

\label{fig:EnergyGrowth}
\end{figure}

Till now, we have obtained the one-on-one correspondence between the
instantaneous energy gain and the moment of inertia (or cloud size).
The initial energy growth as a function of the atomic cloud size is
plotted in Fig.\ref{fig:EnergyGrowth} for several typical power parameters.
The energy gain resulting from the interaction increasing can conveniently
be measured in experiments according to the measurement of the cloud
size in expansion. We find that expanding to the same size, the energy
gain of the system reaches the maximum when the scattering length
increases as $\sim\sqrt{t}$, which provides an identification of
the maximum energy growth theorem in experiments.

\section{Thermodynamic relation\label{sec:Thermodynamic-relation}}

In the follows, we derive an out-of-equilibrium thermodynamic relation
for interacting quantum gases. Here, a two-fluid hydrodynamic theory
is adopted to study the dynamics at nonequilibrium, which is applicable
in a wide range of temperature. The basic idea of the two-fluid hydrodynamic
theory lies in considering the system as if it were a mixture of two
different liquids at finite temperature: a superfluid without viscosity
and a viscous normal fluid. Analogous to those for superfluids or
normal fluids, the number density of particles satisfies the equation
of continuity,
\begin{equation}
\frac{\partial n}{\partial t}+\nabla\cdot{\bf j}=0,\label{eq:7.1}
\end{equation}
where $n=n_{n}+n_{s}$ is the total number density, and ${\bf j}=n_{n}{\bf v}_{n}+n_{s}{\bf v}_{s}$
is the current density with the number density $n_{s(n)}$ for the
superfluid (normal) component and the corresponding velocity ${\bf v}_{s(n)}$.
Besides, the current density ${\bf j}$ satisfies an additional equation
\citep{Landau1987F}
\begin{equation}
\frac{\partial j_{i}}{\partial t}+\sum_{k}\frac{\partial\Pi_{ik}}{\partial x_{k}}+\frac{n}{m}\frac{\partial V_{ext}}{\partial x_{i}}=-\sum_{k}\frac{\partial\Pi_{ik}^{\prime}}{\partial x_{k}}\label{eq:7.2}
\end{equation}
for the $i$th component of ${\bf j}$, where $V_{ext}$ is the external
potential, and $\Pi_{ik}$ is the momentum current density tensor
defined as 
\begin{equation}
\Pi_{ik}\equiv n_{n}v_{ni}v_{nk}+n_{s}v_{si}v_{sk}+\frac{p}{m}\delta_{ik},\label{eq:7.3}
\end{equation}
and $p$ is the pressure. The viscous normal fluid introduces an additional
dissipation described by
\begin{equation}
\Pi_{ik}^{\prime}=-\frac{\eta}{m}\sigma_{ik}-\delta_{ik}\left(\frac{\zeta_{1}}{m}\sigma_{1}^{\prime}+\frac{\zeta_{2}}{m}\sigma_{2}^{\prime}\right),\label{eq:7.4}
\end{equation}
with 
\begin{eqnarray}
\sigma_{ik} & \equiv & \frac{\partial v_{ni}}{\partial x_{k}}+\frac{\partial v_{nk}}{\partial x_{i}}-\frac{2}{3}\delta_{ik}\nabla\cdot{\bf v}_{n},\label{eq:7.5}\\
\sigma_{1}^{\prime} & \equiv & \nabla\cdot\left[mn_{s}\left({\bf v}_{n}-{\bf v}_{s}\right)\right].\label{eq:7.6}\\
\sigma_{2}^{\prime} & \equiv & \nabla\cdot{\bf v}_{n}.\label{eq:7.7}
\end{eqnarray}
Here, $\eta$ and $\zeta_{1,2}$ are respectively the shear and bulk
viscosities. Eqs.(\ref{eq:7.1}) and (\ref{eq:7.2}) forms a set of
basic equations for describing the dynamics of interacting quantum
gases at finite temperature. 

At zero temperature, the system becomes a pure superfluid, and the
number density of the normal component vanishes as well as corresponding
velocities. Then Eq.(\ref{eq:7.2}) becomes
\begin{equation}
\frac{\partial\left(nv_{i}\right)}{\partial t}+\sum_{k}\frac{\partial}{\partial x_{k}}\left(nv_{i}v_{k}\right)+\frac{1}{m}\frac{\partial p}{\partial x_{i}}+\frac{n}{m}\frac{\partial V_{ext}}{\partial x_{i}}=0,\label{eq:7.8}
\end{equation}
which, combining with the equation of continuity, yields
\begin{equation}
\frac{\partial{\bf v}}{\partial t}+\left({\bf v}\cdot\nabla\right){\bf v}+\frac{\nabla p}{mn}+\frac{\nabla V_{ext}}{m}=0.\label{eq:7.9}
\end{equation}
By using the identity of
\begin{equation}
\nabla\left(\frac{v^{2}}{2}\right)={\bf v}\times\left(\nabla\times{\bf v}\right)+\left({\bf v}\cdot\nabla\right){\bf v},\label{eq:7.10}
\end{equation}
and concerning the fact that the superfluid corresponds to a potential
flow with $\nabla\times{\bf v}=0$, we arrive at
\begin{equation}
\frac{\partial{\bf v}}{\partial t}+\nabla\left(\frac{v^{2}}{2}\right)+\frac{\nabla p}{mn}+\frac{\nabla V_{ext}}{m}=0.\label{eq:7.11}
\end{equation}
This is the well-known Euler equation for a superfluid (see for example
in \citep{Pethick2008B}). Here, all the quantities are those for
the superfluid, so that the subscript $s$ is dropped off without
confusion. 

Above the critical temperature, the system is entirely a normal fluid,
and the superfluid portion vanishes. Then Eq.(\ref{eq:7.2}) becomes
the Euler equation for a normal fluid in the presence of viscosity
at finite temperature \citep{Elliott2014O,Landau1987F}, i.e., 
\begin{multline}
\left[nm\frac{\partial}{\partial t}+nm\left({\bf v}\cdot\nabla\right)\right]v_{i}+\frac{\partial p}{\partial x_{i}}+n\frac{\partial V_{ext}}{\partial x_{i}}\\
=\sum_{k}\frac{\partial}{\partial x_{k}}\left(\eta\sigma_{ik}+\delta_{ik}\zeta_{2}\sigma_{2}^{\prime}\right).\label{eq:7.12}
\end{multline}
Consequently, we find that the two-fluid hydrodynamic theory provides
a rather general formulism to describe the dynamics of interacting
many-body systems in a wide range of temperature. Moreover, the quantum
pressure comes into play at zero temperature, which could formally
be included in the definition of pressure, i.e., $p=p_{c}+p_{q}$
and $p_{c\left(q\right)}$ stands for the classical (quantum) pressure
\citep{Manfredi2001S}.

Now we are ready for the discussion of the evolution of moment of
inertia $I\left(t\right)\equiv\left\langle mr^{2}\right\rangle $
based on the two-fluid hydrodynamic formulism. The $i$th component
of $\left\langle r^{2}\right\rangle $ is defined as $\left\langle x_{i}^{2}\right\rangle =\int d{\bf r}nx_{i}^{2}$
with $i=x,y,z$, which obeys
\begin{equation}
\frac{d\left\langle x_{i}^{2}\right\rangle }{dt}=\int d{\bf r}\frac{\partial n}{\partial t}x_{i}^{2}=-\int d{\bf r}\left(\nabla\cdot{\bf j}\right)x_{i}^{2}=2\int d{\bf r}j_{i}x_{i}.\label{eq:7.13}
\end{equation}
By further taking the second-order derivative with respect to $t$,
we have
\begin{equation}
\frac{d^{2}\left\langle x_{i}^{2}\right\rangle }{dt^{2}}=2\int d{\bf r}\frac{\partial j_{i}}{\partial t}x_{i}.\label{eq:7.14}
\end{equation}
According to Eq.(\ref{eq:7.2}) , we obtain
\begin{equation}
\frac{d^{2}\left\langle x_{i}^{2}\right\rangle }{dt^{2}}=2\int d{\bf r}\left(\Pi_{ii}+\Pi_{ii}^{\prime}\right)-2\int d{\bf r}x_{i}\frac{n}{m}\frac{\partial V_{ext}}{\partial x_{i}}.\label{eq:7.15}
\end{equation}
By inserting the explicit form of $\Pi_{ii}^{(\prime)}$ into above
equation and assuming that the external potential is a harmonic trap,
i.e., $V_{ext}\left({\bf r}\right)=m\omega^{2}r^{2}/2$, we arrive
at
\begin{multline}
\frac{1}{4}\frac{d^{2}\left\langle x_{i}^{2}\right\rangle }{dt^{2}}=\int d{\bf r}\left[n_{n}\frac{v_{ni}^{2}}{2}+n_{s}\frac{v_{si}^{2}}{2}+\frac{p}{2m}\right.\\
\left.-\frac{\eta}{2m}\sigma_{ii}-\frac{1}{2m}\left(\zeta_{1}\sigma_{1}^{\prime}+\zeta_{2}\sigma_{2}^{\prime}\right)\right]-\int d{\bf r}\frac{n}{m}\left(\frac{1}{2}m\omega^{2}x_{i}^{2}\right).\label{eq:7.16}
\end{multline}
Summing the above equation over all three directions and noticing
$\sum_{i}\int d{\bf r}\eta\sigma_{ii}=0$, we finally obtain
\begin{multline}
\frac{1}{4}\frac{d^{2}I}{dt^{2}}=E_{kin}-E_{ho}+\frac{3}{2}\mathcal{J}-\frac{3}{2}\int d{\bf r}\left(\zeta_{1}\sigma_{1}^{\prime}+\zeta_{2}\sigma_{2}^{\prime}\right),\label{eq:7.17}
\end{multline}
where $E_{kin}\equiv\int d{\bf r}\left(n_{n}mv_{n}^{2}+n_{s}mv_{s}^{2}\right)/2$
corresponds to the total kinetic energy of the normal fluid and superfluid,
$E_{ho}=\int d{\bf r}n\left(m\omega^{2}r^{2}/2\right)$ is the potential
energy, and $\mathcal{J}\equiv\int d{\bf r}p$ is the integral of
the local pressure over the whole space, namely the thermodynamic
potential. Eq.(\ref{eq:7.17}) simply reduces to that for a normal
fluid with $\zeta_{1}=0$ above the critical temperature \citep{Elliott2014O}.

Combining Eq.(\ref{eq:7.17}) and the dynamic virial theorem, we find
the following simple relation for interacting quantum gases
\begin{equation}
\mathcal{J}=\frac{2}{3}E_{internal}\left(t\right)+\frac{\hbar^{2}\mathcal{C}\left(t\right)}{12\pi ma\left(t\right)}+\Gamma\left(t\right),\label{eq:7.18}
\end{equation}
where we have defined the internal energy $E_{internal}\equiv E\left(t\right)-E_{kin}\left(t\right)-E_{ho}\left(t\right)$
and the dissipation energy resulting from viscosities $\Gamma\left(t\right)=\int d{\bf r}\left(\zeta_{1}\sigma_{1}^{\prime}+\zeta_{2}\sigma_{2}^{\prime}\right)$.
This relation holds for both the normal fluid and superfluid as well
as their mixtures, and thus is valid in a wide range of temperature.
Eq.(\ref{eq:7.18}) is a kind of the out-of-equilibrium analog of
Tan's pressure relation, since it simply reduces to the well-known
form for uniform gases at equilibrium \citep{Tan2008G}, i.e.,
\begin{equation}
p=\frac{2}{3}\mathcal{E}_{internal}+\frac{\hbar^{2}\mathcal{I}}{12\pi ma}\label{eq:7.19}
\end{equation}
with the internal energy density $\mathcal{E}_{internal}$, the contact
density $\mathcal{I}=\mathcal{C}/V$ and the volume $V$.

\section{Conclusions\label{sec:Conclusions}}

It is shown that a variety of out-of-equilibrium dynamics of interacting
many-body systems are elegantly governed by the dynamic virial theorem.
Its applications in several typical nonequilibrium dynamic processes
of cold atoms are presented, in which observable consequences in experiments
are discussed. Remarkably, the dynamic virial theorem provides an
experimentally accessible way to verify the maximum energy growth
theorem according to the measurement of the atomic cloud size in expansion.
Besides, the dynamic virial theorem gives rise to a simple thermodynamic
relation, the analog of Tan's pressure relation at equilibrium, for
interacting many-body systems in the framework of two-fluid hydrodynamic
theory. This thermodynamic relation holds in a wide range of temperature.
Our results provide a fundamental understanding of generic behaviors
of interacting many-body systems at nonequilibrium, and are readily
examined in future experiments with ultracold atoms.
\begin{acknowledgments}
We would like to thank Ran Qi for inspiring discussion on the derivation
of dynamic virial theorem. We are particularly grateful to Xi-Wen
Guan for his helpful suggestions. This work is supported by the National
Natural Science Foundation of China under Grant No.11974384, the National
Key R\&D Program under Grant No. 2022YFA1404102, K. C. Wong Education
Foundation under Grant No. GJTD- 2019-15 and the Natural Science Foundation
of Hubei Province under Grant No. 2021CFA027.
\end{acknowledgments}


\begin{thebibliography}{99}
\bibitem{Derivation1}We may set $y=\dot{b}$, and have $\ddot{b}=\dot{y}=ydy/db$.
The differential equation satisfied by $b$ can easily be solved by
using the initial condition $y\left(b=1\right)=0$.
\end{thebibliography}
\end{document}